\documentclass[12pt]{article}
\let\section=\subsection     \let\subsection=\subsubsection                
\usepackage{graphicx}

\def\Journal#1#2#3#4{{#1} {\bf #2}, #3 (#4)}

\def\NPA{{\em Nucl.~Phys.} A}

\def\PRL{\em Phys.~Rev.~Lett.~}

\newcommand{\mT}[1]{m_{\perp_{#1}}}
\newcommand{\pT}[1]{p_{\perp_{#1}}}

\newcommand{\UNIT}[1]{\mbox{$\,{\rm #1}$}}

\newcommand{\GeV}{\UNIT{GeV}}

\begin{document}
\begin{center}
   {\large \bf ENERGY LOSS OF HIGH $\pT{}$ HADRONS}\\[2mm]
   {\large \bf BY FINAL HADRONIC STATE}\\[5mm]
   K.~GALLMEISTER, C.~GREINER AND Z.~XU\\[5mm]
   {\small \it Institut f\"ur Theoretische Physik, Universit\"at Giessen\\
   Heinrich-Buff-Ring 16, D-35392 Giessen, Germany \\[8mm] }
\end{center}
\begin{abstract}\noindent
In the context of the `jet quenching' phenomena
typically materialization of the jet is assumed to take place in vacuum
outside the reaction zone.
On the other hand quantum mechanical estimates
give a hadronization time on the order of only a few fm/c
for jets materializing into hadrons with transverse momenta
of $\pT{} \leq 10 \, GeV$, which thus should well take place
inside the fireball.
Typical (in-)elastic collisions of these high
$\pT{}$ particles with the bulk of hadrons of the fireball have
a rather low invariant mass and are thus nonperturbative.
An analysis within an opacity expansion in the number of collisions
by means of the FRITIOF collision scheme for various
hadrons will be presented. It shows that late hadronic collisions
can substantially account for the modification
of the high $\pT{} $-spectrum observed for central collisions at RHIC.
\end{abstract}

\section{Introduction: Qualitative estimates}

As one of the very interesting first results, RHIC experiments have
established a significant suppression of moderately high $\pT{}$
hadrons produced in central A+A collisions compared to rescaled
peripheral collisions or rescaled (and extrapolated)
p+p collisions \cite{PHENIX}.
The results are a clear hint for nuclear medium effect(s) at work.

The most popular explanation is the onset of the occurrence
of so called `jet quenching' \cite{Wang}. It is the idea that a
high energetic parton moving through a color dense (and then deconfined) medium
will lose considerable energy most dominantly by gluon radiation, and so
its final fragmentation will give rise to a leading hadron with
considerable lower energy. A recent and much involved calculation
based on the GLV formalism for a finite number of colored collisions
(and including the Cronin effect as well as a slight modification of
the gluon distribution due to shadowing) provide a good agreement
with the data. Still, though, a phenomenological opacity parameter has to
be adjusted \cite{Levai}. It has also been suggested that simple
incoherent scattering in the further partonic evolution
of the fireball should be of relevance \cite{NVC01}. Contrary,
also the proposal was made that the observed spectra for central
collisions show a significant $\mT{}$ scaling being a manifestation
of a direct remnant of the initial gluons which were liberated from
gluon saturated nuclear distribution functions \cite{SKMV01}.
Here, any possible final state interactions which might
or should alter the distributions are completly discarded.

A major concern, and which lead to the present study, is
that it is typically assumed in all these descriptions that
the partons somehow exits the collision region before finally
fragmenting into a `jet' of hadrons. The potential magnitude of the
hadronization time
(or, to be more precise, the time to build up the hadronic
wavefunction) is based on a relativistic and
simple quantum mechanical estimate for
(a) light or (b) heavy quarks \cite{DKMT}
$$
(a) \hspace*{5mm} t_q^{hadr} \, \approx \, E \, R^2  \hspace*{1cm}
\leftrightarrow \hspace*{1cm}
(b) \hspace*{5mm}
t_Q^{hadr} \, \approx \, \frac{E}{m_Q} \, R   \, \, .
$$
Taking (a) with the average radius of the pion $R_\pi \approx 0.5 $ fm
or (b) substituting $m_Q $ by $m_\rho $ for a $\rho $-meson and
taking $R_\rho \approx 0.8 $ fm, one has for the formation time
a crude understanding given as
\begin{equation}
t_F \, \approx \, 1-1.2 \, (E/GeV) * fm/c   \, \, \, .
\end{equation}
Hence, for leading hadrons with
moderately high $\pT{} \le 10 GeV$
original point-like jet-partons have established already a complete
nonperturbative, transversal wavefunction after traveling a distance
in length of
smaller or equal than 10 fm.
Accordingly, the jets should, to a large fraction,
materialize into hadrons still inside the expanding fireball, which has
a transversal size of $R \approx 8fm +0.5* ct $,
where $t$ denotes the local proper time of the expanding system.

The state becoming the leading hadron will interact as a (pre-)meson
with the surrounding low momentum mesons with a cross section of
roughly $\sigma (t) = \sigma_0 *(t/t_F)$ as long as it is not completely
being build up \cite{DKMT}. $\sigma_0 \approx 10-15 mb$
denotes the total cross section of two mesons. The density of hadrons
in the late fireball changes from about  $1$ fm$^{-3}$ to $0.1$ fm$^{-3}$.
The mean free path
of the fast hadron is then estimated to be $\lambda \approx 1-10 $ fm,
meaning a few collisions $L/\lambda = 0,1,2,3 \ldots $. The system
is potentially rather opaque!

Looking at the available energy for an individual collision
(compare Fig.~1),
even for a value of $\pT{}=10\GeV$, one gets a
$\sqrt s <2\GeV$, if the target is a pion at rest, and only with a $\rho $ as
target, one has an invariant mass of above 2.5\GeV{} for particles with a
transverse momentum larger than 3\GeV{}.

\begin{center}
   \includegraphics[height=7cm,angle=-90]{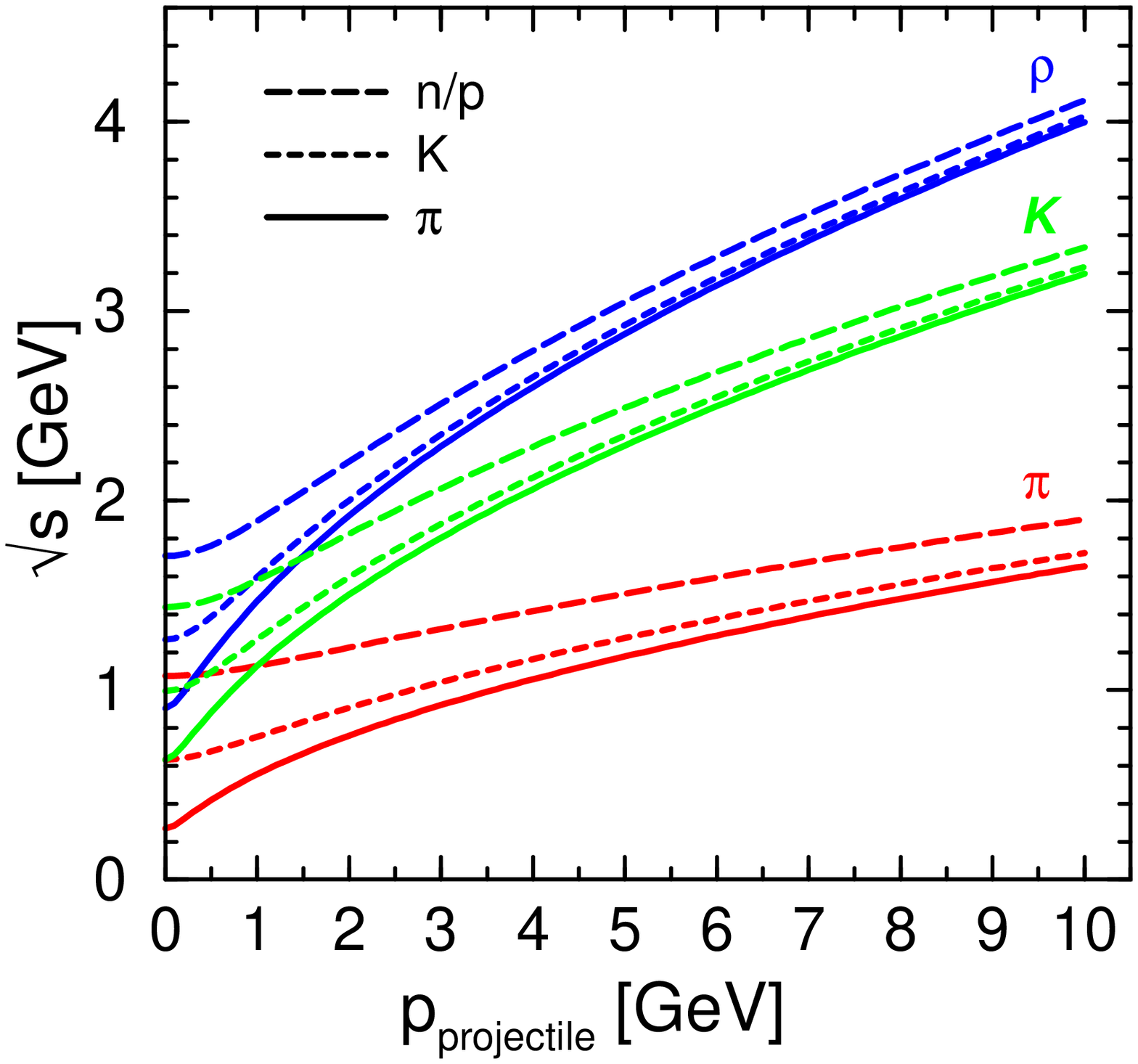}\\
   \parbox{14cm}
        {\footnotesize
        Fig.~1:
        The CMS energy for three targets at rest
        ($\pi,K,\rho$, depicted by the labels near the curves) and
        three types of projectiles ($\pi,K,n/p$, indicated by the solid,
        dashed and long dashed lines) as function of the
        momentum of the projectile.
}
\end{center}

Hence, for the considered transverse
momentum region, one has either elastic scattering, resonance scattering
or also inelastic scattering resulting in a few final hadrons.
In the following we treat the collision with the FRITIOF Monte-Carlo
scheme with a $\rho $-meson as characteristic target hadron being at
rest. To also stress some model dependence, we later compare
with elastic scattering on either a $\rho $
or a $\pi $ as target particles, where the scattering is
taken simply as isotropic \cite{GGX02}.

\section{Energy loss by (multiple) final state hadronic interactions}

The energy loss for one single interaction of a leading hadron {\it i}
with momentum $\pT0 $
on a low transverse momentum target (taken as $\approx 0$, and both hadrons
being located at the same space-time rapidity and momentum
rapidity in a Bjorken-type picture), going into a
hadron {\it j}  with (lower) momentum $\pT{} $ can be written as
\begin{equation}
  f_j(\pT{}) = \sum_i\int d\pT0\ f^0_i(\pT0)\ g_{ij}(\pT0,\pT{})\ ,
\end{equation}
while $f_i^0(\pT0 )$ is the initial distribution, $g_{ij}(\pT0,\pT{})$
is a folding matrix and $f_j(\pT{})$ is the resulting distribution.
$i,j$ stand for the particles $\pi^0$, $\pi^\pm$, $K^0$,
$\overline K^0$, $K^\pm$, $p$, $\overline p$, $n$ and $\overline n$.
The folding matrix $g_{ij}$ depends on the model employed for the
collision (for further details we refer to \cite{GGX02}).

\begin{center}
   \includegraphics[height=8cm,angle=-90]{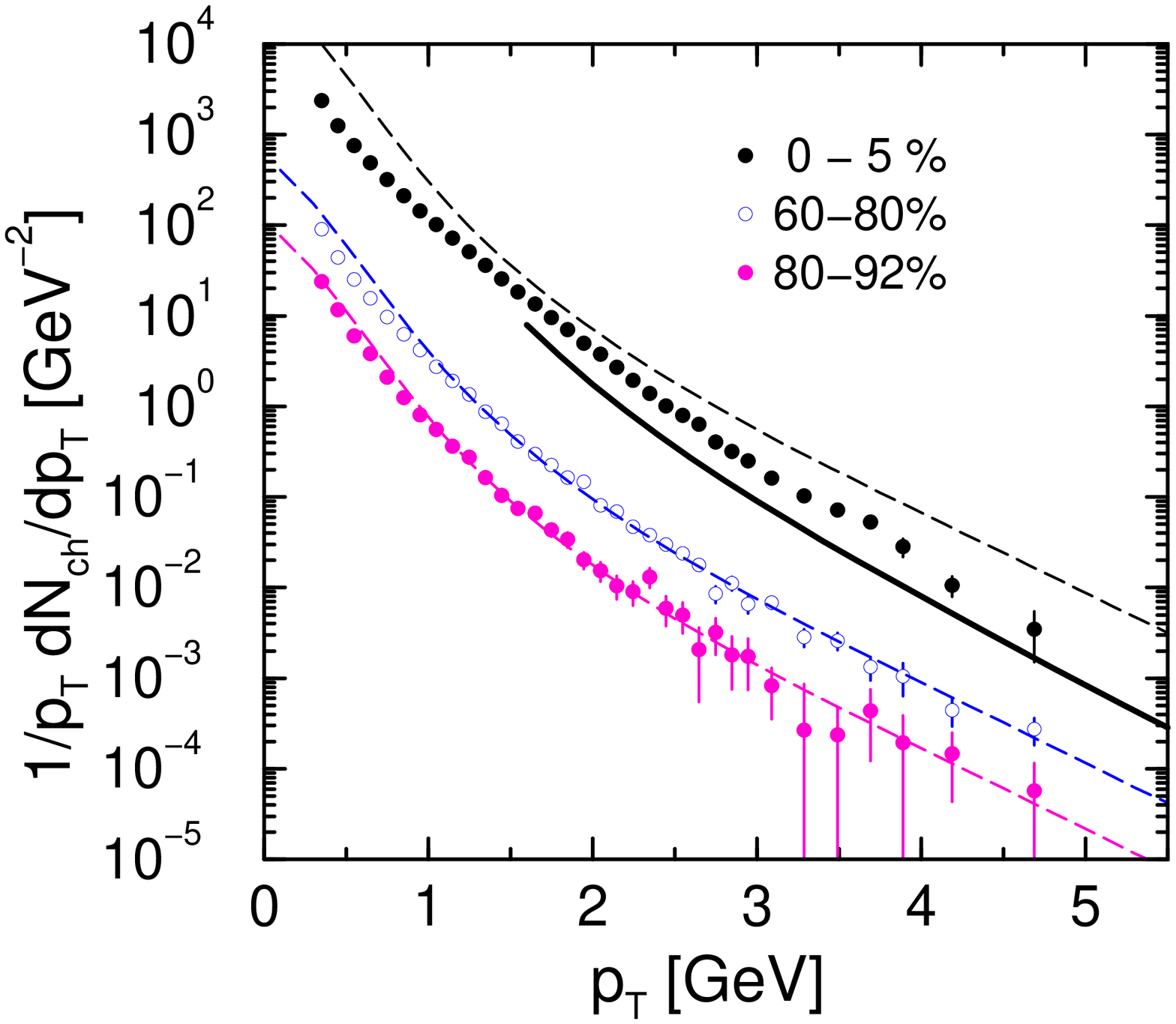}\\
   \parbox{14cm}
        {\footnotesize
        Fig.~2:
      Comparison of the calculations
      at midrapidity (for one unit in rapidity) with the
        (preliminary) PHENIX data \protect\cite{PHENIX2} on
        charged hadrons
        for three different centralities.
        The thin dashed lines indicate the results from PYTHIA,
        scaled with a corresponding binary collision number
        $N_{coll}$.
        The thick solid line depicts the rate for the most central
        region where the particles have suffered on
        average one hadronic collision.
}
\end{center}

For the initial distribution $f_i^0$ of individual hadrons $i$ at
moderately high $\pT{}$ when entering the final hadronic stage
we have no direct information. All various ideas, as mentioned
at the beginning, might contribute and already steepen
the spectrum. In order to see the possible effect of the late hadronic
interactions, we use individual distributions generated by PYTHIA
v6.2 \cite{PYTHIA}.
The parameters were first adjusted to UA1-results of $p\bar{p}$-collisions.
Isospin averaged N-N-collision were then calculated (for details we
refer to \cite{GGX02}).
With the appropriate scaling of binary collisions
($N_{coll}=2.4$ for 80\dots92\%, 12.6 for 60\dots80\% and
945 for 0\dots5\% centrality),
the calculations
are in perfect agreement with the preliminary PHENIX data for charged hadrons
and for neutral pions in the case of peripheral collisions,
as can be seen from Figs~2 and 3.

Moreover, one can see a dramatic effect when employing on average
one single final inelastic collision modelled via the FRITIOF scheme.
In case of the charged hadrons, one single collision
 already would slightly overestimate
the suppression. On the other hand one collision is just right to
perfectly explain quantitatively the modification of the
momentum distribution of the $\pi_0$! (We have only shown
the modifications for $\pT{} > 2 $ GeV, as the spectrum below
should be dominated by soft physics and hydrodynamical expansion.)

\begin{center}
   \includegraphics[height=8cm,angle=-90]{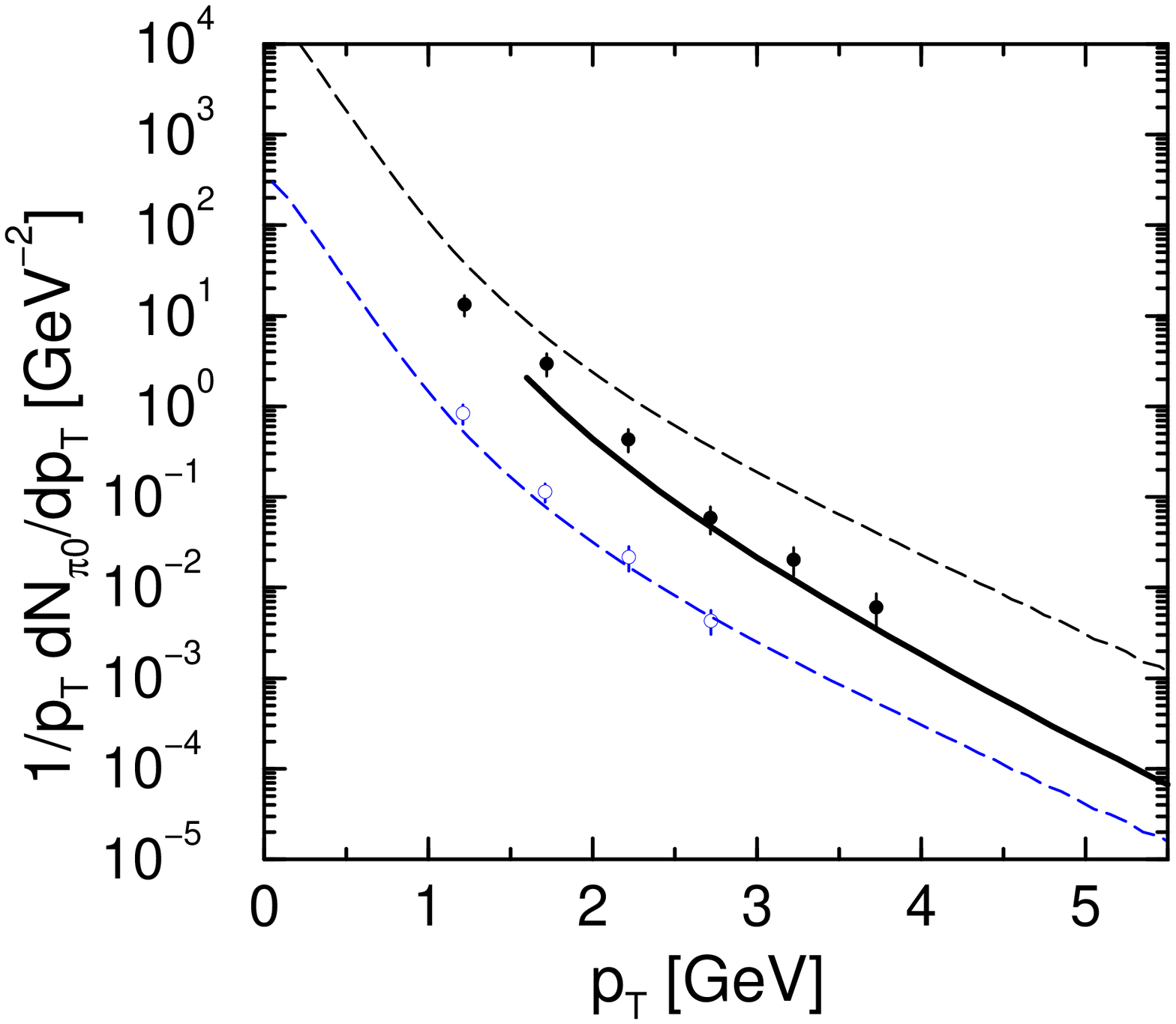}\\
   \parbox{14cm}
        {\footnotesize
        Fig.~3:
       Like Fig.~2, but now a comparison of the present calculations with the
        PHENIX data \protect\cite{PHENIX2} on $\pi _0$.
}
\end{center}

Hence, any potential energy loss by final state hadronic reactions
is in the same range as considered for the jets in deconfined matter.
In Fig.~4 we show calculations for various number of collisions
now at a $\sqrt{s_{NN}}=200 $ GeV up to $\pT{} = 10 $ GeV.
The three employed models (for treating the collisions)
do give more or less similar
predictions. The spectrum steepens considerably
for a larger effective opacity $<L/\lambda>$.
On the other hand at some stage the complete evolution of all
hadronic particles has to be incorporated to see in full
detail the dynamical interplay among `soft' and `hard' hadrons.
The spectrum can not become more steepen than exponential as then
the system has thermalized also in the high $\pT{}$ hadronic
degrees of freedom.

In summary, we have motivated that (pre-)hadrons stemming
from a jet should still materialize in the dense system
for momenta up to 10 GeV. The late hadronic interactions have
a clear and nonvanishing effect in suppressing the spectrum.
On the average one single such interaction should already be enough
to explain quantitatively the RHIC results.
Our finding, however, signals a warning which has to be
addressed in future studies very accurately.
`{\it Omnes viae Romam ducunt}': Various different roads at present
can account for the possible explanation of the modification
in the spectra for central collisions.
Deductions for possible QCD effects of a deconfined
QGP phase on the materializing jets have to disentangle from the
here investigated final state
interactions, before definite conclusions on the importance
of a potential dense  partonic phase, or any other effect, can
quantitatively be drawn.

\begin{center}
   \includegraphics[height=9cm,angle=-90]{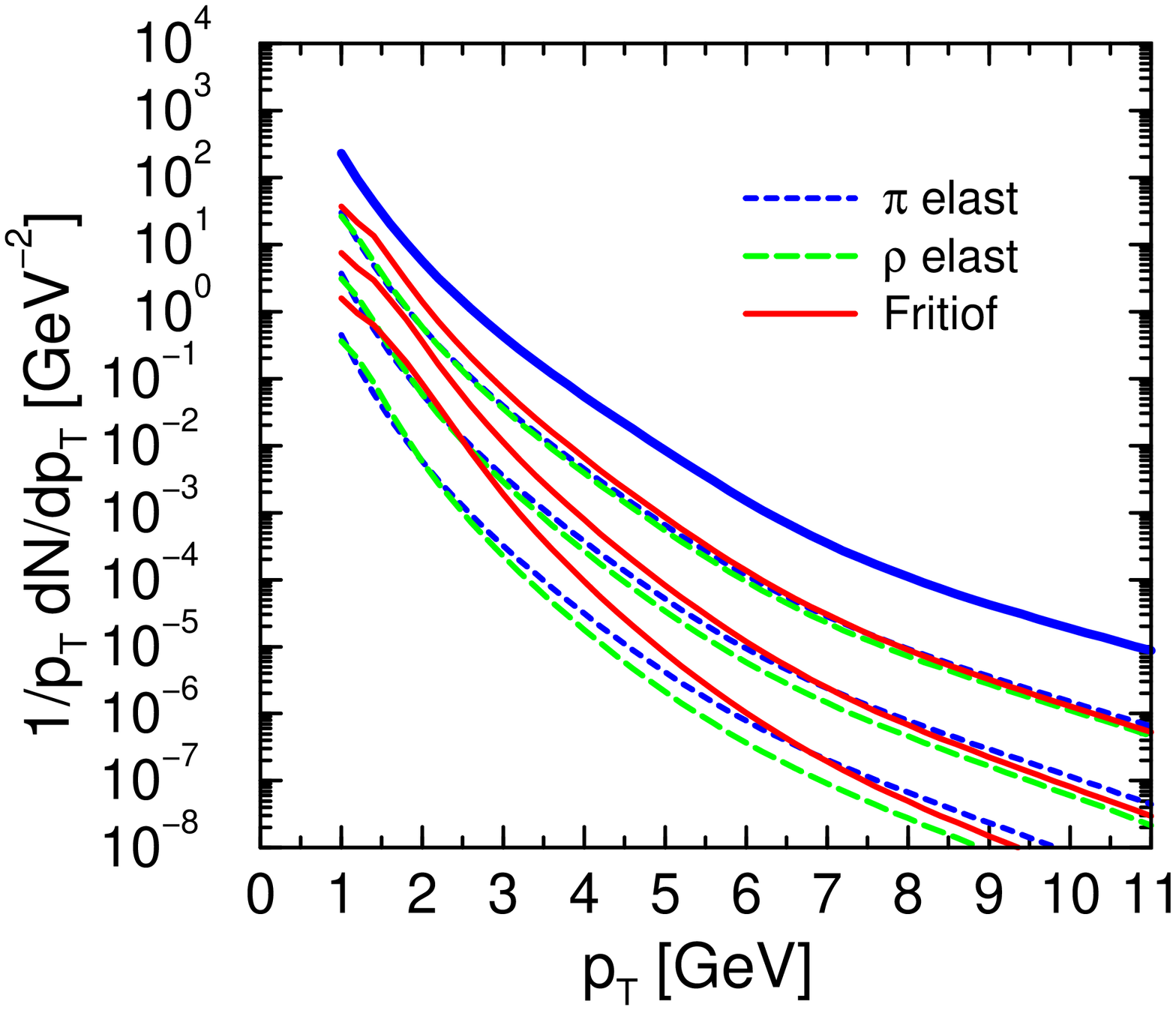}\\
   \parbox{14cm}
        {\footnotesize
        Fig.~4: Resulting $\pT{} $-spectrum of charged hadrons at midrapidity
        for $\sqrt{s_{NN}}=200 $ GeV and for
        $<L/\lambda>\equiv 0,\, 1, \, 2, \, 3$
        collisions. Hadronic collisions are treated in three different ways
        (compare text).
}
\end{center}


\begin{thebibliography}{99}
\itemsep=0cm
\bibitem{PHENIX} K.~Adcox et al [PHENIX Collaboration],
\Journal{\PRL}{88}{022301}{2002}; C.~Adler et al [STAR Collaboration],
\Journal{\PRL}{87}{112303}{2001}; A.~Drees, these proceedings.

\bibitem{Wang} X.N.~Wang, \Journal{\NPA}{698}{296}{2002}.

\bibitem{Levai} P.~Levai, G.~Papp, G.~Fai, M.~Gyulassy, G.~Barnaf\"oldi,
I.~Vitev and Y.~Zhang, \Journal{\NPA}{698}{631}{2002};
P.~Levai, these proceedings.

\bibitem{NVC01} Y.~Nara, S.~Vance and P.~Csizmadia, nucl-th/0109018.

\bibitem{SKMV01} J.~Schaffner-Bielich, D.~Kharzeev, L.~McLerran and
R.~Venugopalan, nucl-th/0108048; J.~Schaffner-Bielich, these proceedings.

\bibitem{DKMT} Y.~Dokshitzer, V.~Khoze, A.~Mueller and S.~Troyan,
`Basics of perturbative QCD', Editions Frontieres (1991).

\bibitem{GGX02} K.~Gallmeister, C.~Greiner and Z.~Xu, in preparation.

\bibitem{PYTHIA} T.~Sj\"ostrand et al, {\em Comp.~Phys.~Commun.} {\bf 135},
238 (2001).

\bibitem{PHENIX2} W.~Zajc et al [PHENIX Collaboration],
\Journal{\NPA}{698}{39}{2002}.
; A.~Drees, these poceedings.




\end{thebibliography}
\end{document}